\documentclass[preprint,superscriptaddress,showpacs]{revtex4}

\usepackage{amsmath}
\usepackage{amstext}
\usepackage{latexsym}
\usepackage[dvips]{graphicx}
\usepackage{hyperref}

\newcommand{\ket}[1]{\left\vert#1\right\rangle}
\newcommand{\bra}[1]{\left\langle#1\right\vert}

\newlength{\defbaselineskip}
\newcommand{\virgo}[1]{``#1''}

\begin{document}

\title{Quantum cloning  in spin networks}
\author{Gabriele De Chiara}
\affiliation{ NEST- INFM \& Scuola Normale Superiore, piazza dei
Cavalieri 7 , I-56126 Pisa, Italy}
\author{Rosario Fazio}
\affiliation{ NEST- INFM \& Scuola Normale Superiore, piazza dei
Cavalieri 7 , I-56126 Pisa, Italy}
\author{Chiara Macchiavello}
\affiliation{INFM \& Dipartimento di Fisica \virgo{A. Volta}, 
Via Bassi 6,I-27100 Pavia, Italy}
\author{Simone Montangero}
\affiliation{ NEST- INFM \& Scuola Normale Superiore, piazza dei
Cavalieri 7 , I-56126 Pisa, Italy}
\author{G. Massimo Palma}
\affiliation{ NEST- INFM \& Dipartimento di Tecnologie
dell'Informazione, Universita' degli studi di Milano\\ via
Bramante 65, I-26013 Crema(CR), Italy}

\date{\today}
\begin{abstract}
We introduce a new approach to quantum cloning based on spin networks and 
we demonstrate that phase covariant cloning can be realized using no 
external control but only with a proper design of the Hamiltonian of the 
system. In the $1 \to 2$ cloning we find that the XY model saturates the 
value for the fidelity of the optimal cloner and gives values comparable to 
it in the general $N \to M$ case. We finally discuss the effect of external 
noise. Our protocol is much more robust to decoherene than a conventional 
procedure based on quantum gates.
\end{abstract}

\maketitle
Quantum information processing protocols~\cite{nielsen00} are
typically described and analysed in terms of qubits and quantum gates, 
i.e. of quantum networks. 
Very recently increasing attention has been devoted to the possibility of
implementing the desired task by a tailored design of a spin network
and an appropriate choice of the couplings and the evolution time. Most 
notably in these schemes the couplings between qubits are fixed and 
this may turn to be an important advantage in the implementation of quantum 
protocols with solid state  devices. Quantum computation for a spin network 
based on Heisenberg couplings was discussed in Ref.~\cite{benjamin03}. 
It was also shown that unmodulated Heisenberg chains can be used to transfer unknown 
quantum states over appreciable distances ($\sim 10^2$  lattice sites) with 
reasonably high fidelity~\cite{bose03,subrahmanyam03}. 
Even perfect transfer could be achieved over arbitrary distances in 
spin chains by a proper choice of the modulation of the coupling 
strengths~\cite{christandl03}, if local measurements on the individual spins 
can be implemented~\cite{verstraete03} or when communicating parties 
have access to limited numbers of qubits in a spin ring~\cite{osborne03}. 
Together with the understanding of the dynamics of entanglement in spin 
systems~\cite{bose03,subrahmanyam03,entdynamics}, this approach to quantum 
communication may lead to the implementation of more complicated protocols 
such as entanglement swapping, teleportation or cloning, just to mention a 
few of them. 

The aim of this Letter is to show that the dynamics of  spin networks with 
fixed couplings can be applied successfully to the problem of quantum cloning. 
The no-cloning theorem~\cite{wootters82} states the
impossibility to make a perfect copy  of an unknown quantum state. 
This no-go theorem has profound implications not only at a
fundamental level but also for practical reasons, since it is the key 
ingredient to guarantee the security of quantum cryptographic
protocols~\cite{gisin02}. Although perfect cloning is forbidden by the laws
of quantum mechanics, it is of great interest to optimize the performance of
approximate cloning machines. In a pioneering work in this direction 
a transformation for copying an unknown qubit state with a state-independent 
fidelity, known as $1\to 2$ universal cloning, was presented~\cite{buzek96}.
It was later proved to be optimal~\cite{bruss98}. The more general problem of  
$N \to M$ universal cloning, where $N$ copies of an unknown input pure 
state are cloned to $M$ output approximate copies, has been also 
addressed~\cite{cloner_n-m}. Notice that the fidelity of the cloning
transformation can be increased if some prior partial knowledge of the states 
to be cloned is available. The first state-dependent cloner was proposed 
in~\cite{bruss98}, where cloning of two non orthogonal states was analysed.
Another interesting example of non universal cloning, which we will consider 
in this paper, is the case of qubits lying on the equator of the Bloch sphere. 
This class of cloners was proposed in Ref.\cite{bruss00} and is known as the 
Phase Covariant Cloning (PCC).

Several protocols for implementing cloning machines have been already 
achieved experimentally~\cite{cummins02,bouwmeester02,demartini,ekert03}. 
In all these proposals to clone a quantum state, the required set 
of operations is realized by means of quantum gates, or otherwise
a post-selection of the state needs to be performed.
(For example, the quantum network corresponding to the $1\to 2$ PCC
consists of two C-NOT gates together with a controlled rotation~\cite{niu99}.)
In the following we concentrate on the phase covariant case and 
we will address the question whether it is possible to perform optimal
cloning without almost any external control. We will show that this is 
indeed possible by choosing a proper Hamiltonian and the topology of 
the spin network. In Fig.1 we give some examples. It turns out  
that it is then possible to clone a quantum state with very high fidelity, 
in some cases with optimal fidelity, by letting the system evolve for a given 
time lapse $t_c$. 
The choice of $t_c$ is the only control which we assume (and need) 
to have on the system. The results of this work are summarized in 
Figs.\ref{fid} and \ref{fidnoise}. We will show that the time $t_c$ 
required for the protocol does not seem to systematically   
increase, for the case considered, as a function of  $N$ and $M$. 
This is a great advantage over the conventional schemes if the  
unavoidable effect of decoherence is taken into account (see 
Fig.\ref{fidnoise}). 

The model Hamiltonian necessary to accomplish our task 
is given by
\begin{equation}
        H_{\lambda} = \frac{1}{4} \sum_{ij} J_{ij}(\sigma^i_x \sigma^j_x + 
        \sigma^i_y \sigma^j_y
        + \lambda \sigma^i_z \sigma^j_z ) + \frac{B}{2} \sum_{i}  \sigma^i_z
\label{eq:hamiltonian}
\end{equation}
where $\sigma^i_{x,y,z}$ are the Pauli matrices, $J_{ij}$ are the 
exchange couplings defined on the links joining the sites $i$ and $j$ and 
$B$ is an externally applied magnetic field. In all the cases we consider 
in this work the couplings $J_{ij}$ are different from zero only if $i,j$ 
are nearest neighbours. To specify which couplings are non-zero one has to 
define the topology of the spin network. The anisotropy parameter $\lambda$
ranges from $0$ (XY Model) to $1$ (Heisenberg Model)~\cite{footnote}.
Once the Hamiltonian in Eq.\eqref{eq:hamiltonian} is specified the cloning 
protocol consists in the following steps: i) the initialization of the network,
where the states to be cloned are stored in $N$ sites and the remaining 
sites are in a blank state; ii) the evolution of the system 
(without any further manipulation) up to the time $t_c$ at which the state has 
been copied in the (initially) blank states.

\begin{figure}
\centering
\includegraphics*[scale=0.4]{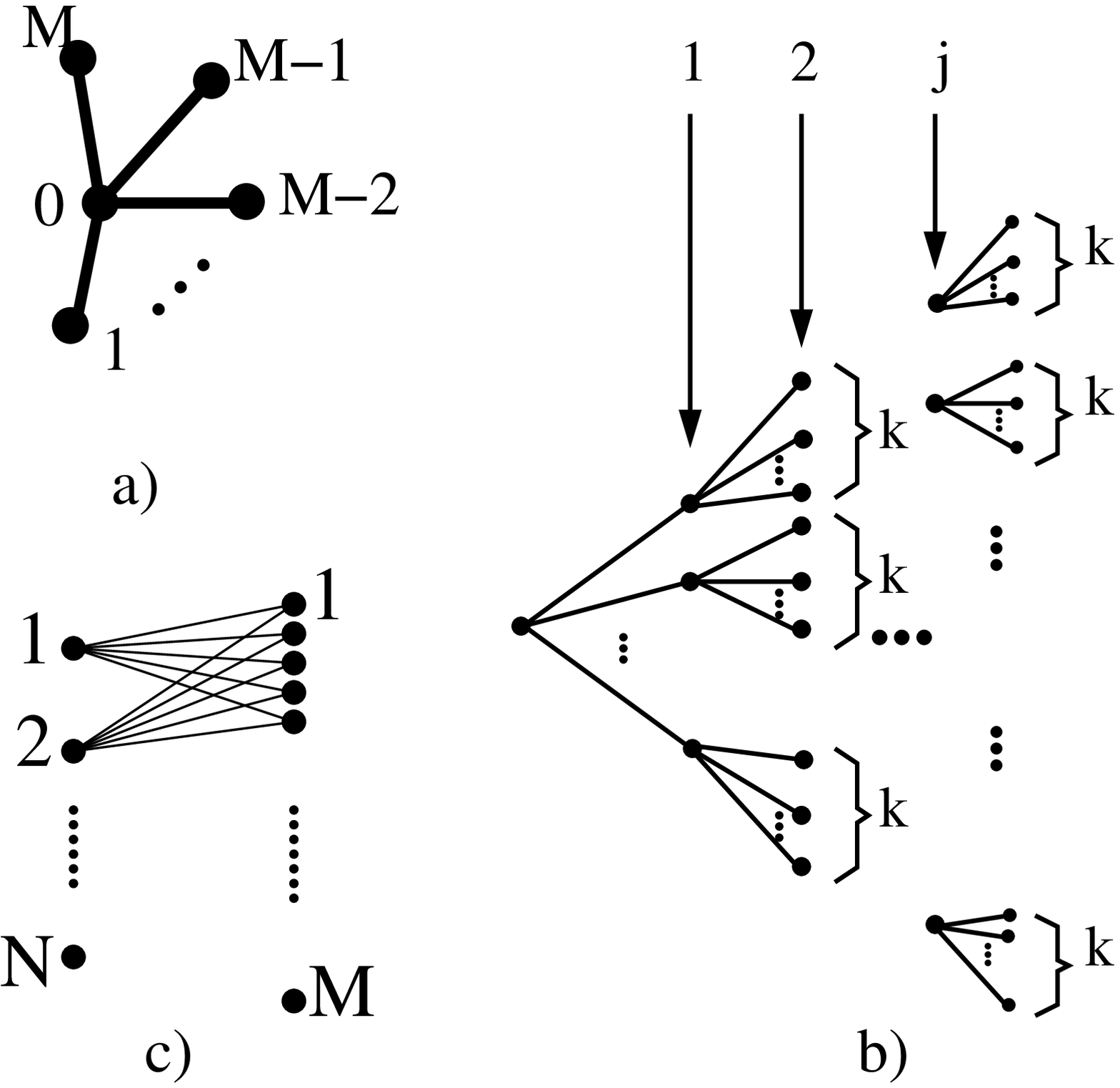}
\caption{Different topologies for $N \to M$ cloner:
 	a)Spin star network for $1 \to M$ cloner. b)Generic graph for the $1
        \to M$ cloner with $j$ intermediate steps and $k$ links
        departing from each vertex. c) Spin network for the $N \to M$ cloning.}
\label{network}
\end{figure}

\begin{figure}
\centering
\includegraphics*[scale=0.3]{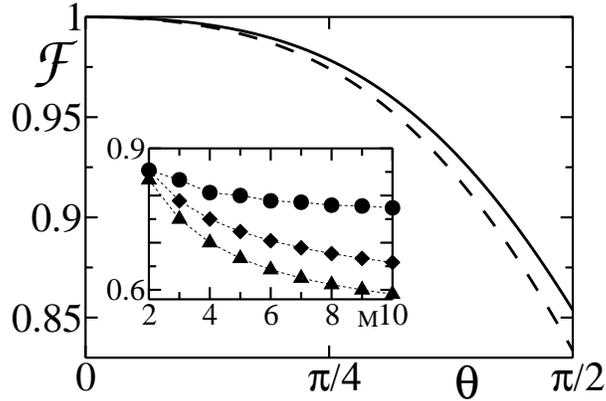}
\caption{The fidelities $\mathcal F_\lambda$ for $M=2$ for generic
    	  $\vartheta$ for the XY (solid) and Heisenberg (dashed) model are
        shown. Notice that the optimal fidelity for the PCC is exactly
        that of the XY model. 
	Inset: the fidelity $\mathcal F$ for the three cases PCC (circle), XY
(diamond) and Heisenberg (triangle) as functions of $M$ for
$\vartheta=\pi/2$.} 
\label{fid}
\end{figure}

\begin{figure}
\centering
\includegraphics*[scale=0.3]{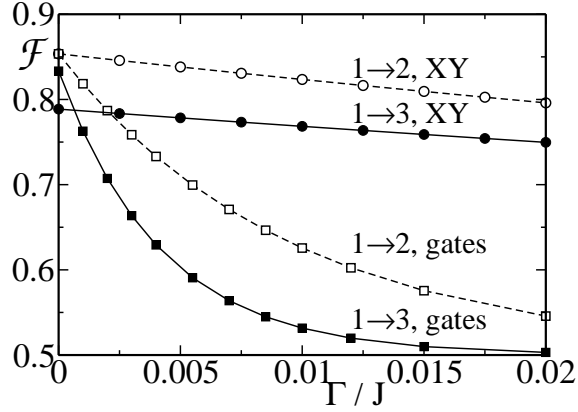}
\caption{Comparison of the fidelity $\mathcal F_\lambda$ obtained by the present 
	method and the quantum circuit discussed in \protect\cite{ekert03,buzek97}
	in the presence of an external noise. White and black symbols refer to 
	the $1 \to 2$ and $1 \to 3$ case respectively ($\vartheta=\pi/2$.)}
\label{fidnoise}
\end{figure}

\emph{\underline{$1\to M$ Cloning - }} A natural choice, although not unique, 
for the graph to be used in this case is a star network as illustrated in
Fig.\ref{network}a.
 The spin in the state to be cloned is placed in the center of the star 
and will be labeled by $0$. The outer spins in the star, labeled from $1$ 
to $M$, are the blank qubits on which the state will be copied. 
Note that the central spin plays a double role of the state to be cloned 
and that of the ancilla. The initial state of the network is
\begin{equation} \label{psitot}
        \ket{\Psi}= \cos\frac{\vartheta}{2} \ket{00\dots 0} + e^{i\varphi}
               \sin\frac{\vartheta}{2}\ket{10\dots 0}
\end{equation} 
($\ket{0}$ and $\ket{1}$ are eigenstates of $\sigma_z$ with
eigenvalues $\pm 1$).

Let us first consider the Heisenberg model ($\lambda =1$ in 
Eq.(\ref{eq:hamiltonian})). We present the results only  for $B=0$, as the
presence of $B$ in this case just changes the time at which the 
maximum fidelity is achieved.
The system can be considered as a spin-1/2 (placed in the central site) 
interacting with with a spin-S particle $S=M/2$ via an exchange interaction
$H_{\lambda=1} = J \vec{S_0}\cdot\vec S$ where $\vec S_0 =
\vec\sigma_0/2$ and   $\vec S=1/2\sum \vec\sigma^i$ is the total spin
of the outer sites.  
The energy eigenstates coincide with those of the total angular momentum 
$\vec I =\vec S_0+\vec S$, labeled as $\ket{I,I_z}$, and can be written 
in the computational basis using Clebsh-Gordan coefficients~\cite{footnote2}. 
The eigenvalues can be found noting that $\vec{S_0} \cdot
\vec{S} = 1/2 (I^2 - S^2 - S_0^2)$. After calculating the total
state of the star at time $t$ we are interested in the reduced density matrix
$\rho_{out}$ of one of the outer spins. Notice that the system is 
invariant under any permutation of the outer spins so their reduced density 
matrices are equal. This property guarantees the symmetry requirement 
of cloning machines, i.e. all the clones are equal. 
We evaluate the quality of the cloning transformation in terms of the fidelity
$F = \bra{\psi}\rho_{out} \ket{\psi}$ of each outer spin with respect to 
the initial state of the qubit to be cloned $\ket{\psi}$.
The fidelity is optimized at times $t^{(M)}_c=\frac{2\pi}{J(M+1)}$ where it takes 
the maximum value $\mathcal F_{\lambda=1} = \mbox{max}_{t}\{F\}$
\begin{equation} \label{fideheisenberg}
        \mathcal{F}_{\lambda=1} = \frac{4\! +\! (3\! +\!
        M)[M\!+\!(M\!-\!1)\cos\vartheta] -  
        (M\!-\!1)\! \cos\! 2\vartheta }{2 {( 1 + M ) }^2} 
\end{equation}

The other interesting case to be considered (which turns out to be the 
optimal as compared with the Heisenberg model) is the XY model ($\lambda =0$ 
in Eq.(\ref{eq:hamiltonian})). We are interested only in the eigenspaces 
with total angular momentum $(M\pm 1)/2$ since the total angular momentum is a
good quantum number. We define the eigenstates of angular momentum of
the outer spins as $\ket{j,m_z}$ where $j$ and $m_z$ are the modulus and
$z$ component of  $\vec S$. In the subspace $j=M/2$ the eigenstates and 
eigenvalues are
$
\ket{\psi^{\pm}_{j,m_z}}=
\frac{1}{\sqrt{2}}\left (\ket{1}\ket{j,m_z}\pm\ket{0}\ket{j,m_z-1}\right )
$
$
E_\pm=\pm \frac{J}{2}\sqrt{(j+m_z)(j-m_z+1)}+B(m_z-\frac{1}{2})  \label{heigenv}
$

for $m_z=j,j-1, \dots ,-j+1$. The other eigenstates are $\ket{0}\ket{j,j}$ 
and $\ket{1}\ket{j,-j}$ with eigenvalues $\pm B(j+1/2)$ 
respectively~\cite{footnote3}. One can show that in the XY model the 
fidelity is maximized for $t^{(M)}_c=\frac{\pi}{\sqrt{M}J}$ and 
$B=\frac{J}{2}\sqrt{M}$ and the maximal fidelity is 
\begin{equation} \label{fidexy}
        \mathcal{F}\!_{\lambda=0}  = \frac{1\! +\! \sqrt{M}\! +\! 2M\! + 
2(M\!-\!1)\cos\!\vartheta\! -\!
        (\sqrt{M}\!-\!1)\cos\! 2\vartheta}{4M}
\end{equation}
In both cases of Eqs.(\ref{fideheisenberg},\ref{fidexy}) the fidelity 
does not depend on $\varphi$. 

An analysis of Eqs.(\ref{fideheisenberg},\ref{fidexy}) allows to 
draw several conclusions along the outlined motivations of this 
work. Let us begin with the comparison of the 
XY versus Heisenberg cloning machines. In Fig.\ref{fid}
we show the results for the maximum fidelity for the XY and Heisenberg models
for the $1 \rightarrow 2$ cloning as a function of $\vartheta$. 

In this case the fidelity for the XY model \emph{coincides} with that of 
the optimal PCC~\cite{bruss00,fiurasek03}. We have thus 
demonstrated that for $M=2$ phase covariant cloning can be realized without 
external gates. Note that for $M=2$ 
and equatorial qubits $\mathcal F_{\lambda=1} =5/6$, i.e. it is 
exactly the value for the universal
quantum cloner~\cite{buzek96,bruss98}. The previous observation, together 
with the fact that the Heisenberg cloner is less accurate, shows that 
the choice of the model is crucial in order to realize the required protocol. 
For generic $M$ and for $\vartheta=\pi/2$ the three fidelities are compared in 
the inset of Fig.\ref{fid}. The optimal PCC fidelity for generic $M$ was derived in 
Ref.\cite{dariano03} and it is reported for comparison in the same
plot.
By increasing $M$ the optimal fidelity scales differently for the two 
models, namely as $1/M$ for the Heisenberg Hamiltonian and as $1/\sqrt{M}$ for 
the XY model.  In both cases the maximum fidelity is lower 
than that of the PCC. However one should note that for both models the time 
at which the maximum fidelity is achieved \emph{decreases} with increasing 
number of copies. As we will show later, our protocol will be much more 
efficient in the (unavoidable) presence of decoherence.

Until now we concentrated our attention to the $1 \to M$ cloning using 
the spin star configuration. Obviously this is not a unique choice and 
one can analyse how  the results presented up to now depend on the topology 
of the spin network.
For the $1 \to M$ cloner we studied a configuration of the spins in tree 
graphs like the one depicted in Fig.\ref{network}b. The initial state is 
placed at the root of the tree while the qubits, onto which the original 
state will be copied, are on the last level. Each graph is characterized by the 
number $k$ of links departing from each site and the number $j$ of intermediate 
levels between the top and the blank qubits level. Our results show that 
the fidelity is almost independent of the graph. For example, by considering 
the XY model and $\vartheta=\pi/2$, $\mathcal F_{\lambda=0}= 0.676$  
for $M=8$ ($k=2$, $j=2$) while for $M=27$ ($k=3$, $j=2$) 
$\mathcal F_{\lambda=0}=0.596$.

We also analysed the role of imperfections in the spin star network
and considered the case in which the coupling
constants are fixed in time but random.
For $J_{1i}$ in the interval $[0.9 J; 1.1 J]$ the average fidelity
(obtained sampling 500 realizations of disorder) decreases only
by less than $0.2\%$ as compared to the ideal cases.

\emph{\underline{$N \to M$ Cloning - }}
In the general case of $N \to M$ cloning and $\vartheta=\pi/2$ we
studied numerically the 
dynamics of the  network shown in Fig.\ref{network}c. We considered
systems up to 9 sites. We concentrated mostly on the case $\lambda=0$ which 
gave higher fidelities compared to the case $\lambda=1$. The results, for 
several values of $N$ and $M$ are shown in the Table~\ref{ntom}.
The values of the fidelity and the times $t_c$ (at which the maximum fidelity is
obtained) confirm the considerations expressed for the $1 \to M$ case. 
The simple cloning procedure we propose is able to attain fidelities which 
are comparable with the optimal fidelities for the $N \to M$ PCC. The time 
$t_c$ in this case is weakly dependent on $N$ and
$M$. This means that the complexity of the protocol does not seem to
increase with increasing number of qubits.

\emph{\underline{Cloning in the presence of Noise - }} The strength of the proposed protocol 
clearly emerges comparing the effect of decoherence on our protocol and on the known cloning quantum circuits \cite{ekert03,buzek97}. We model external
noise as external fields $b_i(t)$ fluctuating independently on each spin of the network. The 
level of the noise is characterized by $\langle b_i(t) b_j(t')\rangle =\Gamma \delta_{ij} \delta (t-t')$ 
(we consider gaussian fluctuations and $\langle b_i(t)\rangle =0$). In 
Fig.\ref{fidnoise} we compare our results  for the $1 \to 2$ and $1 \to 3$ cloning with the
quantum circuits. In order to be consistent we implement
the quantum gates of Refs.\cite{ekert03,buzek97} considering spins interacting through a 
time dependent XY couplings~\cite{schuch03}. Even for a very low noise level $\Gamma /J \sim 10^{-3}$
our protocol is much more efficient than using quantum circuits.
Moreover we believe that in a real implementation also the effective coupling to the environment 
can be different. Our system indeed may be decoupled from the environment during the evolution 
and coupled only at the measurement stage. The circuit model instead is always coupled to the 
environment because an active control on the Hamiltonian is always needed. 

\emph{\underline{Implementation - }}
The networks of Fig.\ref{network} and the XY Hamiltonian can be easily implemented by means
of superconducting nanocircuits~\cite{makhlin01}. Qubits in the charge regime (represented by 
the sites in Fig.\ref{network}) and coupled by Josephson junctions (represented by 
the links in Fig.\ref{network}) realize the optimal PCC described in this work. The preparation 
of the initial state and the measurement procedure can be performed as described, for example,
in Ref.\cite{makhlin01}. Solid state cloning can be realized with present day technology.

This work was supported by the European Community under contracts IST-SQUIBIT, 
IST-SQUBIT2, IST-QUPRODIS, IST-ATESIT, and RTN-Nanoscale Dynamics.

\newpage

\begin{table}
  \begin{tabular}{|c|c|c|c|c|c|}
\hline
 $N$ &$M$ &  $\mathcal F_{PCC}$ &$\mathcal
F$&$J t_c$ &$J/B$ \\
\hline
2&3& 0.941&0.94 &81.04 & 99.8\\
\hline
2&4& 0.933&0.90 &346.75 &49.0\\
\hline
2&5& 0.912& 0.87 &73.66 &95.6\\
\hline
2&6& 0.908&0.83 &277.59 &70.0\\
\hline
2&7& 0.898 & 0.81 &69.04 &17.6\\
\hline
3&4& 0.973 &0.97 &581.07 &17.2\\
\hline
4&5& 0.987 &0.97 &584.65 &57.0\\
\hline
  \end{tabular}
  \caption{The maximum fidelity $\mathcal F$ for $N \to M$ for the
 network of figure \ref{network}c. $\mathcal F_{PCC}$ is the optimal
 fidelity for the PCC \cite{dariano03}. Column 5 (6) reports the
 corresponding evolution time $t_c$ 
  (interaction strength $J$). The results refer to the XY model
 ($\lambda=0$).
The value $\mathcal F$ is found by numerical maximization in the
 intervals $J/B \in [0;100]$ for $N+M<9$ and $J/B \in [0;60]$ for
 $N+M=9$ and $J t \in [0;3 \cdot 10^3]$.}
  \label{ntom}
\end{table}


\begin{thebibliography}{99}
\bibitem{footnote}
        The model is invariant under rotations around the $z$
        axis so it is a natural candidate for a PCC.
        We also analysed the case in which the XY symmetry is broken, 
        the results for the fidelity 
        were always much worse than for XY or Heisenberg models.
\bibitem{footnote2}
        To find the evolution for the initial state, Eq.\eqref{psitot}, 
        we need to
        express it in the eigenstate basis. After some simple algebra one
        finds that 
        $ \ket{00\dots 0} = \ket{\frac{M+1}{2},\frac{M+1}{2}}$ and
        $\ket{10\dots 0}  = \frac{1}{\sqrt{1+M}}\ket{\frac{M+1}{2},
        \frac{M-1}{2}}
        -\frac{\sqrt{M}}{\sqrt{1+M}}\ket{\frac{M-1}{2},\frac{M-1}{2}}$.
\bibitem{footnote3}
        The initial state of the star is the same as for the Heisenberg model
        \protect{\eqref{psitot}} and it can be written in terms of eigenstates 
        of the  Hamiltonian as 
        $\ket{\Psi(0)} = \cos\frac{\vartheta}{2} \ket{0}\ket{j,j}  + 2^{-1/2} 
        e^{i \varphi}\sin\frac{\vartheta}{2}[
        \ket{\psi^+_{M/2,M/2}}+\ket{\psi^-_{M/2,M/2} }]$.
\end{thebibliography}
\end{document}